\documentclass[12pt,preprint]{aastex}

\begin{document}

\title{Benzene formation in the inner regions of protostellar disks}
\shorttitle{Benzene formation in protostellar disks}


\author{Paul M. Woods and Karen Willacy} \affil{Jet Propulsion
Laboratory, California Institute of Technology,\\ MS 169-506, 4800 Oak
Grove Drive, Pasadena, California 91109, USA.\email{Paul.M.Woods@jpl.nasa.gov}}
\shortauthors{Woods and Willacy}

\begin{abstract}
Benzene (\textit{c-}C$_6$H$_6$) formation in the inner 3\,AU of a
protostellar disk can be efficient, resulting in high abundances of
benzene in the midplane region. The formation mechanism is different
to that found in interstellar clouds and in protoplanetary nebulae,
and proceeds mainly through the reaction between allene (C$_3$H$_4$)
and its ion. This has implications for PAH formation, in that some
fraction of PAHs seen in the solar system could be native rather than
inherited from the interstellar medium.
\end{abstract}


\section{Introduction}

Benzene is the smallest aromatic molecule, and is presumed to be a
basic building block in the formation of polycyclic aromatic
hydrocarbons (PAHs). PAHs have been observed in protostellar disks
around both low-mass T Tauri \citep{gee06} and intermediate-mass
Herbig Ae/Be (HAeBe) stars \citep{ack04}. The number of detections in
T Tauri-type disks is low, although this is to be expected since the
incident UV radiation field is several orders of magnitude smaller
than in HAeBe stars. Models indicate that the PAH emission comes from
surface layers \citep{hab04}, leading to speculation as to whether
PAHs form in these high UV environments or elsewhere.

Both PAHs and benzene are also seen in protoplanetary nebulae (PPNe)
-- PAHs in the circumstellar tori and in the bipolar outflows
\citep{mat04}, and benzene in the torus around \object{CRL618}
\citep{cer01}. However, the processes that form benzene depend on the
environment, with different pathways dominating in the interstellar
medium (ISM), in PPNe, in planetary atmospheres \citep{leb05} and, as
we shall show, in protostellar disks. In this Letter we consider the
formation of benzene in a protostellar disk, and the implications this
has for PAH formation.

\section{Benzene formation in different environments}\label{sec:diffenvs}

The gas-phase synthesis of benzene proceeds via different reaction
pathways in different environments, depending on physical and chemical
conditions. One pervading condition for high abundances of benzene is
that of high density ($\gtrsim$10$^{9}$\,cm$^{-3}$). Several reaction
schemes have been suggested for benzene formation:

\paragraph{The interstellar medium}
A reaction scheme for the gas-phase formation of benzene was proposed
by \citet{mce99} following the measurement and calculation of new
rates for radiative association reactions. The formation pathway for
benzene they suggested is thus:
\begin{eqnarray}
\rm C_4H_2^+ + H &\longrightarrow&\rm C_4H_3^+ + h\nu \label{c4h3form}\\
\rm C_4H_3^+ + C_2H_2 &\longrightarrow&\rm {\mathit c-}C_6H_5^+ + h\nu \label{c6h5form}\\
\rm {\mathit c-}C_6H_5^+ + H_2 &\longrightarrow&\rm {\mathit c-}C_6H_7^+ + h\nu \\
\rm {\mathit c-}C_6H_7^+ + {\mathit e^-} &\longrightarrow&\rm {\mathit c-}C_6H_6 + H \label{c6h6form}
\end{eqnarray}
and variations of the above (for instance, C$_2$H$_3$ may replace
C$_2$H$_2$ in reaction~\ref{c6h5form}). This produces moderate amounts
($\sim$10$^{-9}$, with respect to H$_2$) of benzene in a
dense (10$^4$\,cm$^{-3}$) interstellar cloud model.

\paragraph{Protoplanetary nebulae}
\citet{woo02,woo03} applied the above reaction scheme to the dense
torus ($n\sim$10$^{9}$\,cm$^{-3}$) around the protoplanetary nebula
\object{CRL618}. \object{CRL618} is rich in hydrocarbons, and benzene
has been detected in infra-red spectra of this object
\citep{cer01}. In the highly ionising environment around
\object{CRL618} C$_4$H$_3^+$ is formed more efficiently via reactions
of various ions with acetylene:
\begin{eqnarray}
\rm HCO^+ + C_2H_2 &\longrightarrow&\rm C_2H_3^+ + CO\\
\rm C_2H_3^+ + C_2H_2 &\longrightarrow&\rm C_4H_3^+ + H_2 \\
\rm C_2H_2^+ + C_2H_2 &\longrightarrow&\rm C_4H_3^+ + H,
\end{eqnarray}
followed by reactions~\ref{c6h5form}--\ref{c6h6form}. The models of
\citet{woo02,woo03} produced a fractional abundance of benzene of
$\sim$10$^{-6}$, in good agreement with observations.


\section{Model}

\begin{figure}
\plottwo{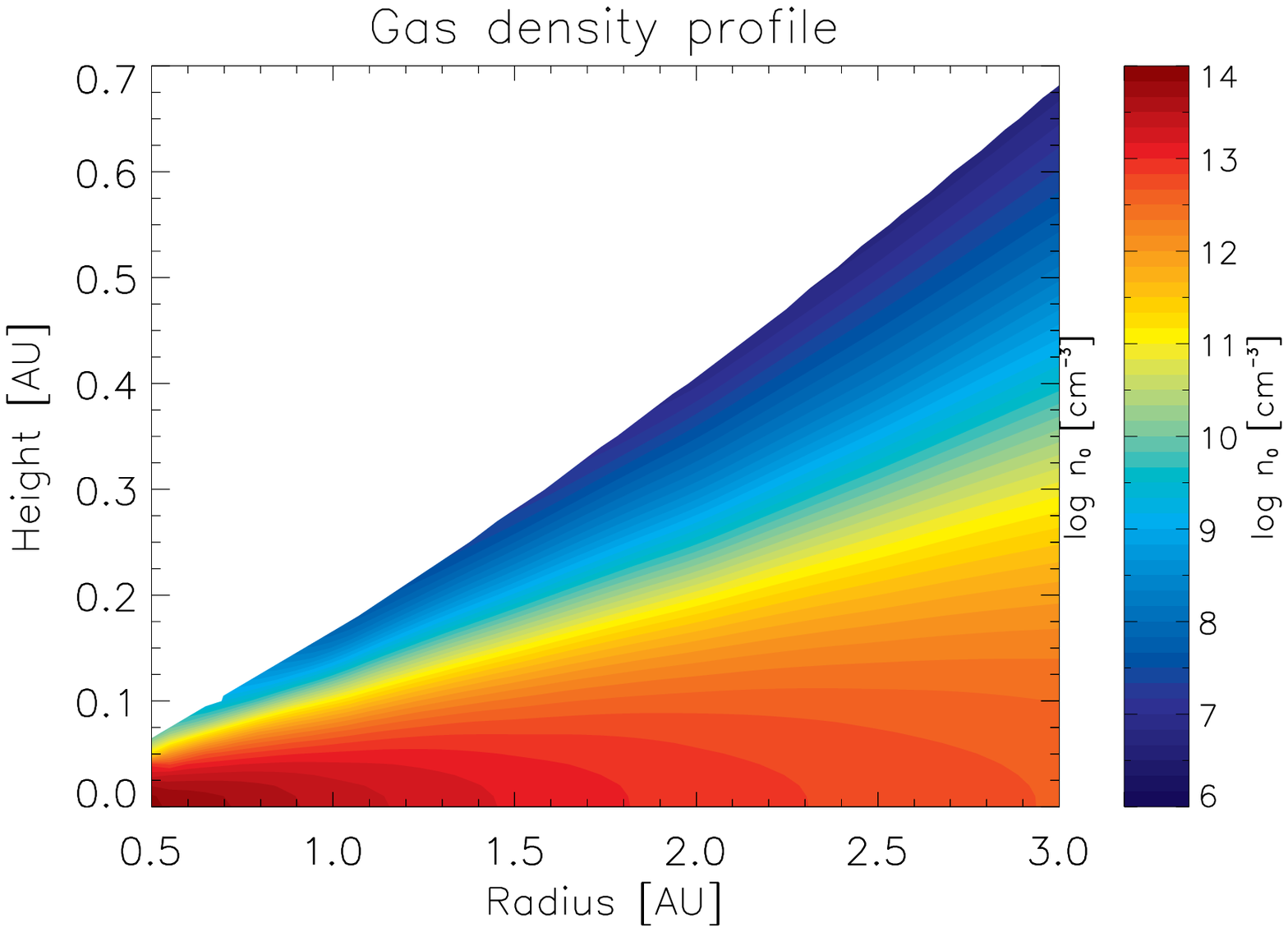}{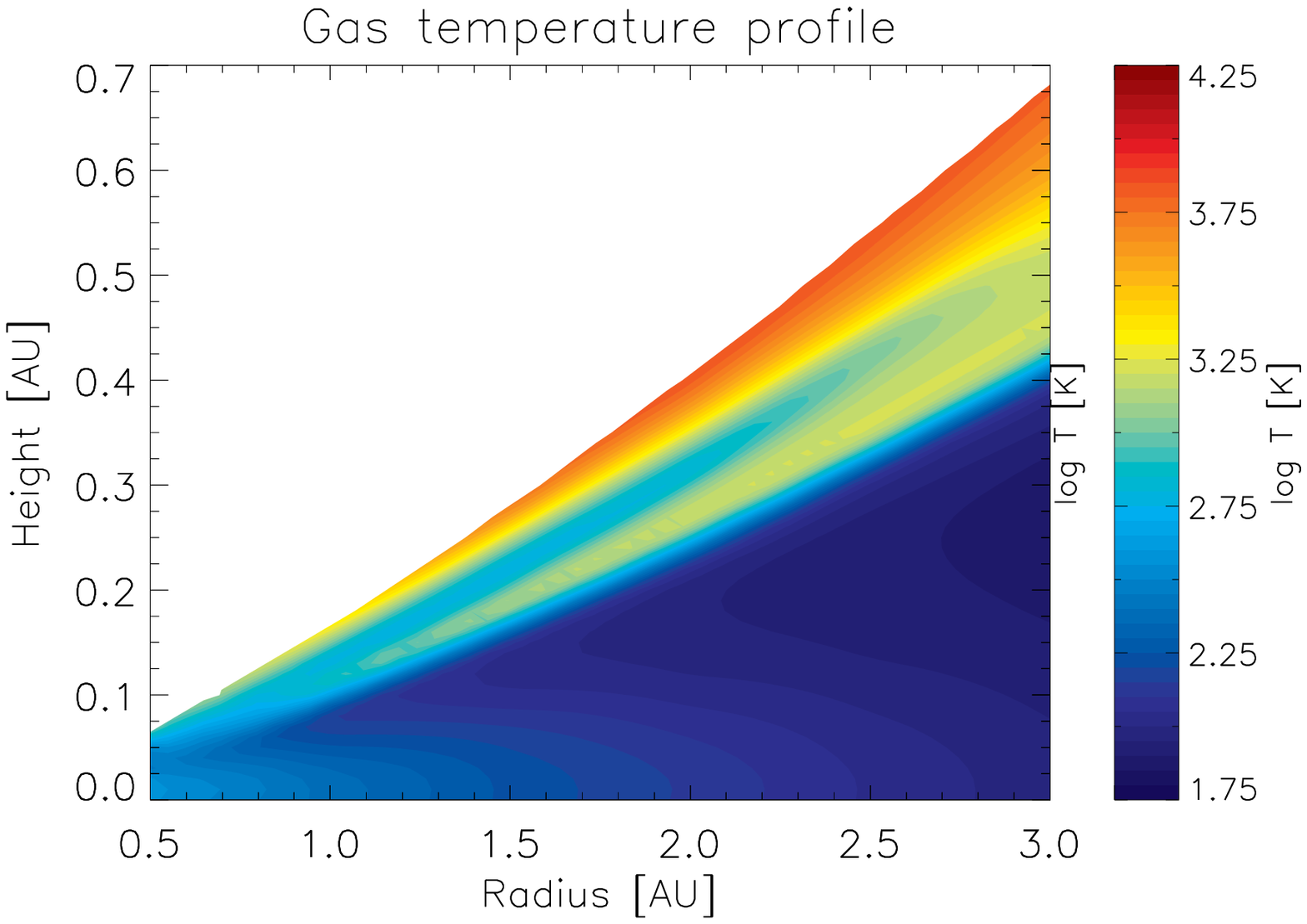}
\caption{Density and gas temperature profiles for the inner 3\,AU of
the disk model. The left panel shows gas density, whereas the right
panel shows the gas temperature calculated by the heating/cooling
balance. See the electronic edition of the Journal for a colour
version of this figure.\label{fig:denstemp}}
\end{figure}

The protostellar disk model we use will be described more fully in a
forthcoming publication, \citet{woo06}. We use a density and dust
temperature profile based on that of \citet{dal01}
(Fig.~\ref{fig:denstemp}). This is a flared accretion disk model, with
a maximum dust grain size of 0.25\,$\mu$m, similar to the size of
interstellar dust. The disk has a mass accretion rate of
$\dot{M}$=10$^{-8}$\,M$_\odot$\,yr$^{-1}$ and a surface density of
$\Sigma$=100\,g\,cm$^{-2}$ at 1\,AU, whilst the central star has the
following properties: $M_\star$=0.7\,M$_\odot$, $T_\star$=4\,000\,K
and $R_\star$=2\,R$_\odot$.  Using these profiles, we calculate UV
photon fluxes throughout the disk with the ray-tracing component of
the model of \citet{yor99}. This calculates the UV field due to the
central star and the interstellar radiation field (ISRF) and also
includes radiation scattering effects. At 10\,AU we assume that the UV
field due to the central star is 50,000 times the ISRF, in accordance
with the observations of \citet{ber03}. Despite the differing spectral
shapes of the ISRF and a typical T Tauri stellar field, with strong
emission lines dominating the T Tauri spectrum \citep{ber03}, the
formation of benzene in the very dense and well-shielded region we
consider is not critically dependent on our choice of stellar UV
field.

Given gas densities, dust temperatures and UV photon fluxes at each
point on our grid\footnote{We use a 35\,AU$\times$16\,AU
(radial$\times$vertical) grid with spacings of 0.5\,AU$\times$0.02\,AU
to model the very inner region of the disk.}, we are then able to
solve the heating and cooling balance of the gas in a similar manner
to \citet{kam01}, \citet{kam04} and \citet{gor04}. The gas temperature
can be underestimated by up to two orders of magnitude if it is
assumed to be equal to the dust temperature throughout the disk,
although this is less critical for the optically thick midplane region
where benzene forms. Figure~\ref{fig:denstemp} shows the range of gas
temperatures within the inner 3\,AU of the disk (the benzene formation
region).

We use a subset of the extensive UMIST Rate99 gas-phase chemical
network \citep{let00}, augmenting it with gas-grain interactions
(freezeout, thermal desorption) and grain surface reactions so that
our reaction set comprises approximately 2400 reactions amongst 200
species. We include species with more carbon atoms than benzene to
ensure that chain-lengthening does not result in a spurious build-up
of benzene. We have selected reactions which are valid in the region
10--500\,K; this upper limit is exceeded in the upper regions of the
disk where temperatures can reach $\sim$10,000\,K, although at such
high temperatures molecules are destroyed. Hence we concentrate on
the portions of the disk closer to the midplane, less than 2
scaleheights.

The model follows the passage of a parcel of gas as it flows inwards
from the outer regions of the disk ($>$35\,AU), and as such it traces
a period in the evolutionary history of the disk. At the outer edge of
the disk the parcel has the composition of a million-year old
molecular cloud. The parcel then accretes onto the central star over a
period of approximately 200\,000 years. As discussed by \citet{wil98},
the choice of initial abundances has relevatively little effect on the
chemistry of most species.

\section{Results}

We find that benzene formation in a protostellar disk is efficient
within 3\,AU of the central star, and fractional abundances of
$\sim$10$^{-6}$ are achieved inside a radius of 2\,AU
(Fig.~\ref{fig:benzdist}). The most efficient formation mechanism for
benzene in this region is due to the reaction between C$_3$H$_4$ and
its ion, C$_3$H$_4^+$, for which the rate has been measured to within
$\pm$25\% in the laboratory
\citep[$k$=7.48$\times$10$^{-10}$\,cm$^3$\,s$^{-1}$;][]{ani84}\footnote{The
UMIST reaction network does not make a distinction in this case
between isotopomers of C$_3$H$_4$; however, the experimental data for
reaction~(\ref{eq:benzform}) was obtained using allene, rather than
methylacetylene (CH$_3$CCH) or cyclopropene (\textit{c-}C$_3$H$_4$)}.
This is followed by neutralisation of the resulting
\textit{c-}C$_6$H$_7^+$ ion in collisions with dust grains:
\begin{eqnarray}
\rm C_3H_4 + C_3H_4^+ &\longrightarrow&\rm \mathit{c-}C_6H_7^+ + H \label{eq:benzform}\\
\rm \mathit{c-}C_6H_7^+ + grain &\longrightarrow&\rm \mathit{g-}C_6H_6 + \mathit{g-}H,
\end{eqnarray}
where \textit{g--} signifies a species adsorbed onto a grain
surface.

In this reaction scheme, the adsorption energy of benzene becomes an
important factor in determining the amount of gaseous
benzene. Estimates of this vary wildly: comparison of similar species
in Table~4 of \citet{has93} leads to a value of 4\,730\,K, which
agrees well with the experiment of \citet{arn88} using a graphite
surface (4\,752\,K). If one takes the approach of \citet{gar06}, and
adds the adsorption energy of acetylene to four times the adsorption
energy of carbon and four times the adsorption energy of hydrogen, one
arrives at the value of 7\,587\,K. This figure lies in the upper range
of an experiment on a graphite surface by
\citet[][4\,840--7\,840\,K]{loz95}. To quantify the effects of these
different adsorption energies, we ran one model with a low binding
energy of 4\,750\,K, and one model with a higher binding energy of
7\,580\,K. Results show a significant decrease in the extent of the
gas-phase benzene distribution with the higher binding energy, as to
be expected, with little change in the peak fractional abundance.

The inner, midplane regions of our disk model are very dense (reaching
$\sim$10$^{14}$\,cm$^{-3}$), possibly dense enough for three-body
(termolecular) reactions to have an effect on the chemistry. In order
to investigate the effect on benzene production in particular, we ran
a third model including a representative three-body reaction:
\begin{equation}
\rm C_3H_3 + C_3H_3 + M \longrightarrow\rm C_6H_6 + M,
\end{equation}
where two propargyl radicals collide with M, a third molecule. This
has been shown to be a very efficient formation mechanism for benzene
when densities are high; see the discussion of \citet{che92}. We have
adopted a rate coefficient from chemical models of the Saturnian
atmosphere:
\begin{equation}
k_\mathrm{ter} = \frac{k_\infty k_0 n_0}{k_\infty n_0 + k_0}\quad\rm cm^3\,s^{-1},
\end{equation}
\citep{mos00} where k$_\infty$, the high pressure limit to the
reaction rate, was taken to be
1.66$\times$10$^{-13}$\,cm$^6$\,s$^{-1}$ and k$_0$, the low pressure
limit, 1$\times$10$^{-27}$\,cm$^3$\,s$^{-1}$
\citep{won00}. Inclusion of this reaction in the model did not result
in a significant change in the abundance of gas-phase benzene, despite
the presence of large fractional abundances of C$_3$H$_3$ in dense
regions. Even adopting the faster rates for k$_\infty$ and k$_0$ from
\citet{leb05} (1.2$\times$10$^{-10}$\,cm$^3$\,s$^{-1}$ and
6.0$\times$10$^{-28}\mathrm{e}^{1680/T}$\,cm$^6$\,s$^{-1}$, respectively) had
little overall effect.

\section{Discussion}

\begin{figure}
\plottwo{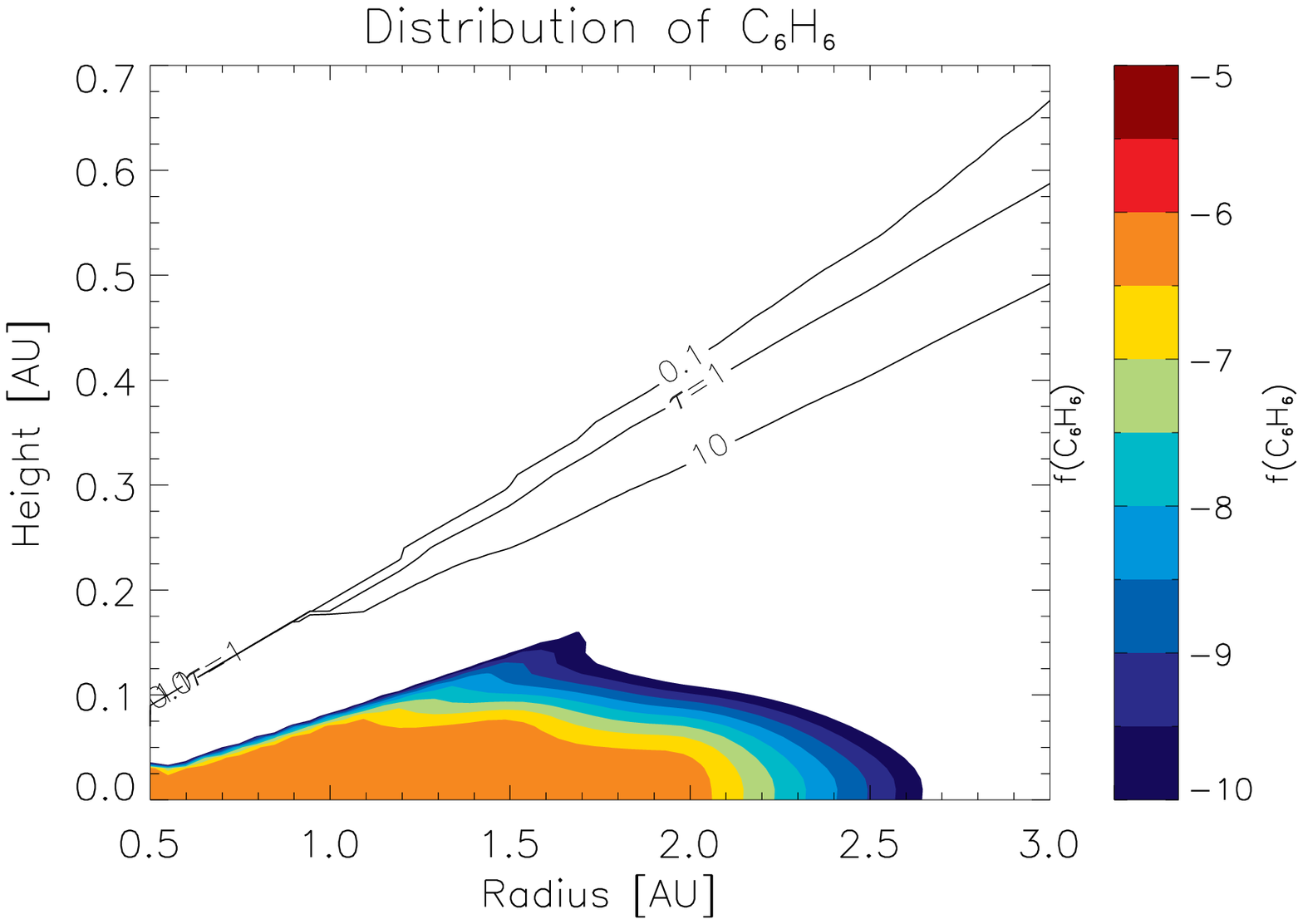}{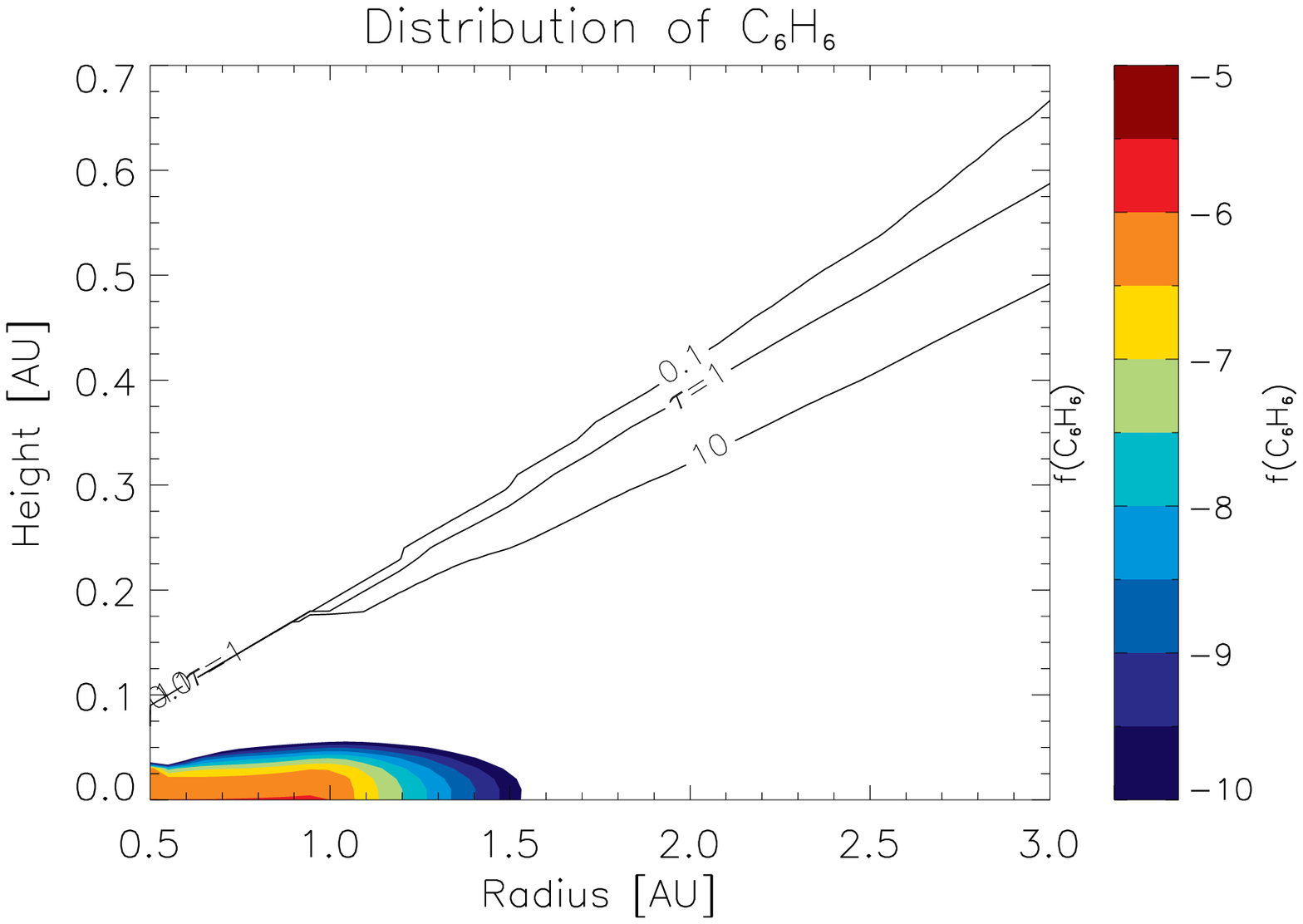}
\caption{The inner 3\,AU of the disk, showing the distribution of
(gaseous) benzene in terms of fractional abundance.  The black solid
lines are optical depth ($\tau$) contours. The left panel shows a
model with a low binding energy for benzene, 4\,750\,K. The right
panel shows the effect of a higher binding energy, 7\,580\,K. See the
electronic edition of the Journal for a colour version of this
figure.\label{fig:benzdist}}
\end{figure}

The region in which benzene is abundant in our model
(Fig.~\ref{fig:benzdist}) is also rich in other organic molecules:
high densities mean fast collision timescales with other molecules and
with dust grains, and also protection from the strong stellar UV field
in this region. These prime conditions for the growth of organic
molecules may allow molecules to grow beyond the single aromatic ring
of benzene into PAHs, either by substitution of hydrogen by acetylene
in the ring \citep{won00,mos00}, or by the dimerisation of benzene
(A$_1$) with successive phenyl rings to produce naphthalene (A$_2$),
phenanthrene (A$_3$), pyrene (A$_4$) and successively large PAH
molecules \citep{che96}. Benzene itself is unlikely to persist through
to the formation of planets: calculations show that benzene has a
timescale for UV destruction in the diffuse solar system of only
hundreds of years \citep{all96,rui05}, meaning that the benzene seen
on Jupiter, Saturn and Titan is unlikely to be primordial.

In order for PAHs to form, the constituents must be present in a
region warm enough for ring closure to occur. PAHs only seem to form
efficiently at temperatures in the region of 900--1\,100\,K
\citep{fre89} and not greater. Production may be possible, too, at
lower temperatures (700--900\,K) given a high pressure
\citep{hel96}. The region of benzene production is at a temperature of
a few hundred degrees, and sits just below a region of sufficiently
high temperature for PAH formation. Invoking vertical mixing here
could raise benzene and other organics into the higher temperature
region, as well as push newly-formed PAHs into the surface layers of
the disk, where they are observed. Mixing timescales at a few AU are
fast enough \citep[$\sim$625\,yr;][]{ilg06} to dredge material from
the midplane to surface layers within the lifetime of a T Tauri disk
($\sim$10$^6$\,yr).

For PAHs to subsist in the solar system, the rate of growth of the PAH
must be greater than the rate of photodissociation. However, once a
PAH reaches 30-40 carbon atoms in size, it is effectively safe from
photodissociation since at this size the infrared radiative rate of
the PAH dominates the photodissociation rate. In the ISM, a benzene
molecule is destroyed (by removal of an acetylene molecule) at a rate
of 1.5$\times$10$^{-10}$\,s$^{-1}$, giving a lifetime of around 200
years. The growth time (accretion of an acetylene molecule) is given
by
$\tau_\mathrm{gr} = (n_0 X_\mathrm{C_2H_2} k_\mathrm{C_2H_2})^{-1}~\mathrm{s}$
\citep{all96}. Inserting typical values for the abundance of acetylene
in the inner part of the disk, $X_\mathrm{C_2H_2}$=10$^{-7}$, and the
reaction rate for accretion of acetylene,
$k_\mathrm{C_2H_2}$=10$^{-11}$\,s$^{-1}$, gives a necessary density of
1.5$\times$10$^8$\,cm$^{-3}$ for PAH growth from benzene. This tallies
with the low abundances of benzene in (low density) interstellar cloud
models (see Sect.~\ref{sec:diffenvs}) and the lack of a detection of
benzene in the ISM. However, these conditions \textit{are} met within
the inner regions of a protostellar disk, as has been shown in our
chemical model (which also includes destruction of benzene by
reactions with ions). In fact, the strength of the UV field drops
below the interstellar level for much of the midplane region, and will
remain at that level for much of the transition towards a debris
disk. Assuming a PAH of 40 carbon atoms forms solely through the
addition of acetylene to a benzene ring, a PAH could be produced from
a benzene molecule on a timescale of approximately fifty years at a
density of 10$^{10}$\,cm$^{-3}$. Even though the actual route to PAH
formation is likely to be different, this estimate does give a
representative timescale and show that PAH production from benzene
could possibly be prolific in this type of environment. We will
consider the formation of larger aromatics in a future paper.

Having ascertained that benzene can survive for an amount of time long
enough for PAHs to form from it, the question arises, do PAHs form in
similar regions to which benzene forms? The answer to this is unclear,
since the supporting observational evidence is sparse. Of the three
detections of PAHs in T Tauri-type disks \citep{gee06}, all three
disks are thought to have inner dust holes (from SED models). Thus it
may just be a selection effect that PAH emission is only detected from
the upper layers of disks because of the presence of these dust holes:
otherwise the continuum emission swamps any possible PAH emission
feature originating from deeper in the disk. Further observations are
necessary to confirm the correlation between the presence of benzene
and PAHs.

\acknowledgments This research was conducted at the Jet Propulsion
Laboratory, California Institute of Technology under contract with the
National Aeronautics and Space Administration. Support comes from an
appointment to the NASA Postdoctoral Program at JPL, administered by
ORAU through a contract with NASA.

\clearpage

\end{document}